\documentclass[preprint,aps,showpacs,showkeys,pre]{revtex4}

\usepackage{graphicx}

\newcommand{\fig}[1]{Fig.~\ref{#1}}

\newcommand{\eq}[1]{Eq.~(\ref{#1})}

\def\siml{\mathrel{\mathchoice {\vcenter{\offinterlineskip\halign{\hfil
$\displaystyle##$\hfil\cr<\cr\sim\cr}}}
{\vcenter{\offinterlineskip\halign{\hfil$\textstyle##$\hfil\cr
<\cr\sim\cr}}}
{\vcenter{\offinterlineskip\halign{\hfil$\scriptstyle##$\hfil\cr
<\cr\sim\cr}}}
{\vcenter{\offinterlineskip\halign{\hfil$\scriptscriptstyle##$\hfil\cr
<\cr\sim\cr}}}}}

\begin{document}


\title{Asymmetric simple exclusion processes with diffusive
  bottlenecks} 
\author{Stefan Klumpp} 
\email{klumpp@mpikg-golm.mpg.de}
\author{Reinhard Lipowsky}
\homepage{http://www.mpikg-golm.mpg.de/lipowsky}
\affiliation{Max-Planck-Institut f\"ur Kolloid- und
  Grenzfl\"achenforschung, 14424 Potsdam-Golm, Germany}

\date{\today}

\begin{abstract}
  One-dimensional asymmetric simple exclusion processes (ASEPs) which
  are coupled to external reservoirs via diffusive transport are
  studied. These ASEPs consist of active compartments characterized by
  directed movements of the particles and diffusive compartments in
  which the particles undergo unbiased diffusion. Phase diagrams are
  obtained by a self-consistent mean field approach and by Monte Carlo
  simulations.  The diffusive compartments act as diffusive
  bottlenecks if the velocity of the driven compartments or ASEPs is
  sufficiently large.  A diffusive bottleneck at the boundary of the
  system leads to the absence of a maximal current phase, while a
  diffusive bottleneck in the interior of the system leads to a new
  phase characterized by different densities in the two active
  compartments adjacent to the diffusive one and to a maximal current
  defined by the bottleneck.
\end{abstract}



\maketitle

\section{Introduction}

Asymmetric simple exclusion processes (ASEPs) are simple
one-dimensional driven lattice gases with hard core exclusion. They
were originally introduced in the context of protein synthesis
\cite{MacDonald__Pipkin1968} and have attracted much interest during
the last years as simple models for boundary-induced
phase transitions
\cite{Krug1991}, for which many rigorous results have been obtained
\cite{Derrida__Mukamel1992,Schuetz_Domany1993,Derrida__Pasquier1993}. 
In the open system,
different stationary states are found, which depend on the rates of
injection and extraction of particles at the ends.  Varying the
injection and extraction rates, or equivalently the densities at the
left and right boundary, both continuous and discontinuous phase
transitions are observed.  The actual stationary state is selected
via the dynamics of domain walls and density fluctuations
\cite{Kolomeisky__Straley1998}.

Promising candidates for the experimental observation of these phase
transitions are systems of cytoskeletal motors which move
unidirectionally along cytoskeletal filaments
\cite{Lipowsky__Nieuwenhuizen2001,Nieuwenhuizen__Lipowsky2002,Klumpp_Lipowsky2003}.
However, these motors unbind from their track after a few seconds,
since their binding energy can be overcome by thermal fluctuations.
Observed over sufficiently long times which exceed a few seconds, they
alternate between the bound and the unbound state and perform peculiar
random walks.  If a motor is bound to a filament, it moves in a
directed way along the filament, while unbound motors diffuse freely.
As motors are strongly attracted by the filament, the motor density
along the filament can be large even if the overall motor
concentration is rather small, which implies that hard core exclusion
plays an important role in the bound state. To study these combined
movements, we have recently introduced a class of lattice models where
bound and unbound motor movements are described as biased and
symmetric random walks on a lattice, respectively
\cite{Lipowsky__Nieuwenhuizen2001,Nieuwenhuizen__Lipowsky2002,Klumpp_Lipowsky2003}.  In
these models, the traffic of motors along a filament is an asymmetric
simple exclusion process with the additional property that motors can
attach to and detach from the track.

For open tube systems with a single filament and fixed motor
concentrations at the tube ends, the same types of phases are found as
for the usual one-dimensional ASEP \cite{Klumpp_Lipowsky2003}.  If the
filament is shorter than the tube and motors have to diffuse over a
certain distance to reach one end of the filament from the left
boundary and again to reach the right boundary from the other end of
the filament, the phase boundaries of the system can be shifted by
changing geometrical tube parameters or motor parameters.
In particular, a maximal current phase, in which the current attains
its maximally possible value, can only be found if the diffusive
currents from the left end of the tube to the filament and from the
filament to the right end of the tube can be as large as this maximal current. 
These diffusive currents however are restricted by the diffusion
coefficient of the unbound motors and by geometric parameters
\cite{Klumpp_Lipowsky2003}.

\begin{figure}[h]
  \includegraphics[angle=0,width=.6\textwidth]{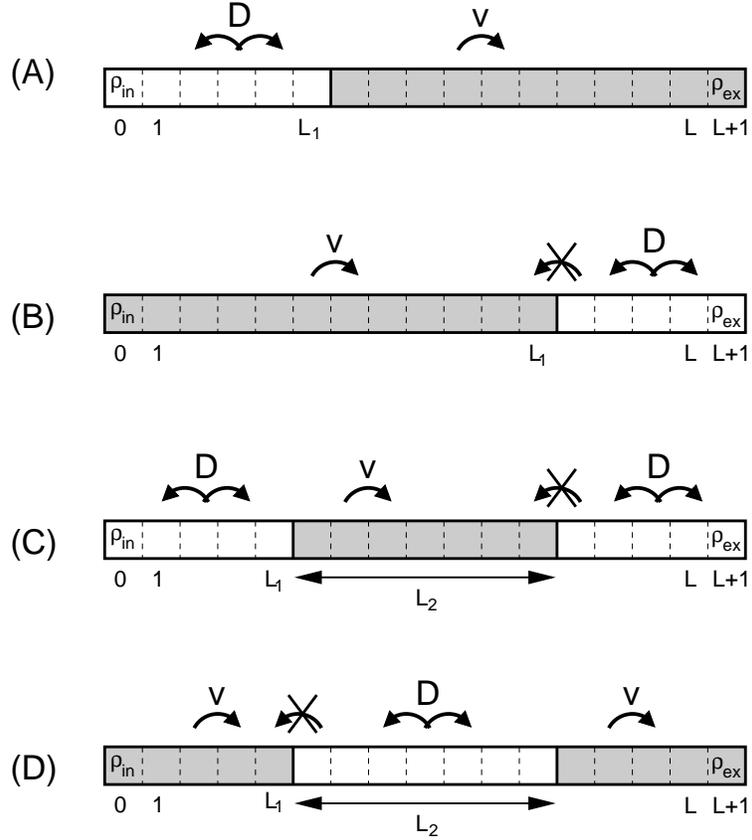}
    \caption{Four geometries of one-dimensional lattices which 
      consist of active compartments (gray) where transport is
      described by an asymmetric simple exclusion process and
      diffusive compartments (white) where transport is described by a
      symmetric exclusion process. The linear extensions of the
      compartments are denoted by $L_1$, $L_2$, and $L_3$.  The total
      length of the system is given by $L=L_1+L_2$ in cases (A) and
      (B) and by $L=L_1+L_2+L_3$ in cases (C) and (D). In the active
      compartments, motion is completely biased and particles hop to
      the right with probability $v$, while in the diffusive
      compartments, particles hop both to the left and to the right
      with probability $D$. In addition, we do not allow particles to
      enter the active compartments from the right. The densities at
      the boundary sites $x=0$ and $x=L+1$ are fixed to $\rho_{\rm
        in}$ and $\rho_{\rm ex}$, respectively.  }
    \label{fig:Bildchen_asep_diff_inj_extr}
\end{figure}

The latter phenomenon is generic and not restricted to the specific
tube geometry. In the following, we study several one-dimensional
systems which consist of compartments characterized by active or
diffusive transport. We will discuss four simple geometries as shown
schematically in \fig{fig:Bildchen_asep_diff_inj_extr}. While
particles move only to the right in the active compartments, forward
and backward steps occur with the same probability in the diffusive
compartments.  For these models, we determine the phase diagram
analytically using a mean field approach to calculate effective
boundary densities or effective injection and extraction rates for the
active compartments. The method is based on the constraint that the
stationary current must be equal in all compartments.

The article is organized as follows. After introducing the model in
section \ref{sec:model}, we discuss {\em diffusive injection and extraction}
of particles into/from an active compartment in section
\ref{sec:diffInjExtr}. We start with only diffusive injection in
section \ref{sec:diff_inj} which corresponds to case (A) in
\fig{fig:Bildchen_asep_diff_inj_extr}, proceed with only diffusive
extraction in
section \ref{sec:diff_extr}, see case (B) in \fig{fig:Bildchen_asep_diff_inj_extr}, and then study 
the case (C), for which particles are both injected and extracted via
diffusive compartments, see section \ref{sec:diff_inj&extr}. We
compare the mean field results with Monte Carlo simulation in section
\ref{sec:simulations}. Finally, we discuss the case of a {\em diffusive
compartment between two active compartments} as shown as case (D) in
\fig{fig:Bildchen_asep_diff_inj_extr} in section \ref{sec:caseD}.
This case corresponds to a defect that must be overcome by unbiased
diffusion.

\section{The model}
\label{sec:model}

In the following, we will discuss transport on one-dimensional
lattices. The coordinate along the lattice is denoted by $x$ and will
be measured in units of the lattice constant $\ell$.

We consider systems that can be decomposed into two or three different
compartments in which transport is either diffusive or directed. The
four cases that will be discussed in the following are shown
schematically in \fig{fig:Bildchen_asep_diff_inj_extr}. The total
length of the system is taken to be $L$ in all cases. The linear
extensions of the compartments are denoted by $L_1$, $L_2$, and $L_3$,
compare \fig{fig:Bildchen_asep_diff_inj_extr}.
In case (A), the system consists of a left compartment with $1\leq
x<L_1$, where transport is diffusive, and a right compartment with
$L_1+1\leq x\leq L_1+L_2=L$, where transport is active or directed. In
case (B), transport is directed in the left compartment 
and diffusive in the right compartment.
The cases (C) and (D) correspond to situations where
the systems consist of three compartments with $L=L_1+L_2+L_3$. In case (C), transport is
driven in the middle compartment with $L_1+1\leq x\leq L_1+L_2$ and
diffusive in the left and right ones. 
Finally in case
(D), transport is directed in two compartments, the left and the right
one, 
but diffusive in the middle compartment.
In all cases, we will assume that the
extensions of the active compartments are sufficiently large, so that
we can neglect finite-size effects.

In the following, we will take the active transport to be always
directed to the right and to be totally asymmetric, i.e., we do not
allow backward steps in the compartments with active transport. At
lattice sites which belong to such an active compartment, particles
attempt to hop to the adjacent lattice site to their right with a
certain probability per unit time $\tau$. We denote this probability
by $v$ since it is equal to the velocity of a particle in the active
compartment (and in the absence of other particles), measured in units of
$\ell/\tau$. The hopping attempt is rejected if the target site is
already occupied by another particle.  In summary, motion in the
active compartment is described by a totally asymmetric simple
exclusion process.

In contrast, motion in the diffusive compartments is described by a
symmetric exclusion process. A particle at a lattice site which
belongs to a diffusive compartment attempts to make a forward and a
backward step with equal probability $D$, which corresponds to the
diffusion coefficient measured in units of $\ell^2/\tau$. Note that we
could eliminate one parameter by measuring time in units of the time
scale for diffusive steps of size $\ell$ by choosing $\tau=\ell/(2D)$
(for this choice, the diffusion coefficient measured in units of
$\ell^2/\tau$ would be given by $D=1/2$). This implies that the results
which we derive in the following will depend only on the ratio $v/D$.
All hopping attempts in the diffusive compartments are again rejected
if the target site is occupied by another particle.  In order to
simplify the following calculations, we do not allow particles to
enter the active compartments from the right, i.e.\ all hopping
attempts from the first site of a diffusive compartment to the last
site of the active compartment to its right --- from $L_1+1$ to $L_1$ in
case (B) and (D) and from $L_1+L_2+1$ to $L_1+L_2$ in case (C) --- are
rejected.

Finally, the densities at the boundary sites $x=0$ and $x=L+1$ have
the fixed values
\begin{equation}
  \rho(x=0)=\rho_{\rm in}\qquad{\rm and}\qquad \rho(x=L+1)=\rho_{\rm ex}.
\end{equation}
These sites are taken to have the same dynamics as the adjacent
compartments of the system. Particles thus attempt to enter the system
from the left with probability $D\rho_{\rm in}$ if the first
compartment is diffusive and with probability $v\rho_{\rm in}$ if it
is an active compartment. Particles at the last lattice site with
$x=L$ leave the system to the right with probability $v(1-\rho_{\rm
  ex})$ if the site $x=L$ belongs to an active compartment and with
probability $D(1-\rho_{\rm ex})$ if it belongs to a diffusive one. In
the latter case, particles also try to enter the system from the right
at $x=L$ with probability $D\rho_{\rm ex}$. Likewise, particles at
site $x=1$ can leave the system with probability $D(1-\rho_{\rm in})$
if $x=1$ belongs to a diffusive compartment.  As before, particles can
only enter the system at $x=1$ or $x=L$ if these sites are not
occupied.

\section{Diffusive injection and extraction of particles}
\label{sec:diffInjExtr}

In this section, we consider the cases (A)--(C), where transport is
driven or active in one compartment, but particles are diffusively
injected and/or extracted from the system and have to diffuse over a
certain distance before they reach the active compartment and/or
before they can leave the system at the right end.

As mentioned before, the active compartment is described by an
asymmetric simple exclusion process. Let us therefore briefly
summarize what is known about the phase diagram of this process, see,
e.g., Ref.\ \cite{Kolomeisky__Straley1998}. In an open system, where
the densities are fixed to $\rho_{\rm in}$ and $\rho_{\rm ex}$ at the
left and right boundary of the ASEP, respectively, the stationary
state is determined by the boundary densities. The stationary states
are characterized by the bulk density $\rho^0$ and the stationary
current $J$.

\begin{figure}[h]
  \begin{center}
    \leavevmode
    \includegraphics[angle=-90,width=.5\textwidth]{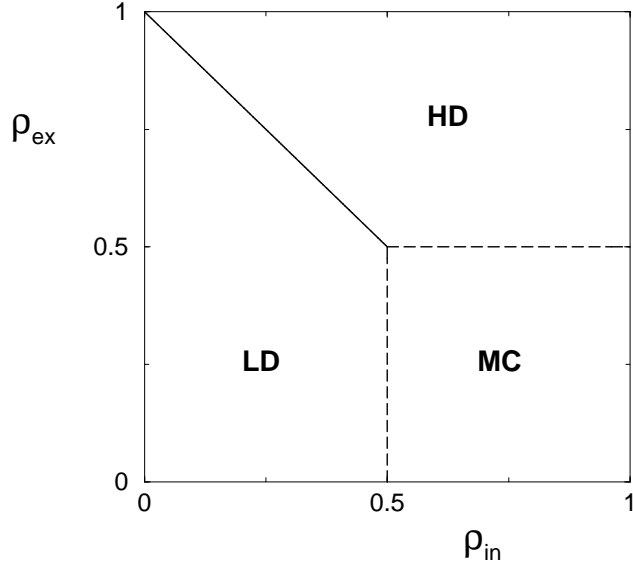}
    \caption{Phase diagram of the usual asymmetric simple exclusion 
      process (ASEP), which describes a single active compartment, as a 
      function of the left
      and right boundary densities $\rho_{\rm in}$ and $\rho_{\rm
        ex}$. }
    \label{fig:phasendiag_normASEP}
  \end{center}
\end{figure}

For the ASEP with open boundaries, three different phases can be 
distinguished as shown in 
\fig{fig:phasendiag_normASEP}: If the density at the
left boundary is relatively small and satisfies $\rho_{\rm in}<1/2$
and $\rho_{\rm in}<1-\rho_{\rm ex}$, the system is in the low density
phase (LD) for which the bulk density is equal to the left boundary
density and the current is $J=v\rho_{\rm in}(1-\rho_{\rm in})$. If the
density at the right boundary is relatively large with $\rho_{\rm
  ex}>1/2$ and $\rho_{\rm ex}>1-\rho_{\rm in}$, the system is in the
high density phase (HD), the bulk density is equal to the right
boundary density, and the current is $J=v\rho_{\rm ex}(1-\rho_{\rm
  ex})$. At the transition from the low density to the high density
phase, the current is continuous, but the bulk density is
discontinuous. Finally, for $\rho_{\rm in}>1/2$ and $\rho_{\rm
  ex}<1/2$, the system is in the maximal current phase, where the
current is maximal, $J=v/4$ and the bulk density is $1/2$. The
transitions to the maximal current phase are continuous.

In contrast to the simple ASEPs just described, our systems are
characterized by the property that at least one of the boundary
densities of the active compartments is not fixed, but adjusted by the
dynamics of the system.  In the following, we will determine the phase
diagrams of these systems using a mean field approach. 
We proceed in three steps and consider first diffusive injection and
extraction of particles separately, combining them in the last step.

\subsection{Diffusive injection of particles}
\label{sec:diff_inj}

First we consider case (A), a system with only diffusive injection of
particles.  Particles leave the system at the right boundary with rate
$v(1-\rho_{\rm ex})$ and no particles enter the system at the right
end.  In the stationary state, the current $J$ must be the same in
both compartments. In the left compartment with $1\leq x\leq L_1$
where transport is diffusive, the density is then given by
$\rho(x)=\rho_{\rm in}-xJ/D$.  Within mean field approximation, the
right compartment with $L_1+1\leq x\leq L$, corresponds to the usual
ASEP with the effective left boundary density
\begin{equation}
  \rho_{\rm in,eff}=\frac{D \rho(L_1)}{v}=\frac{D\rho_{\rm in}}{v}-\frac{L_1 J}{v}.
\end{equation}
as follows from $v\rho_{\rm in,eff}\equiv D \rho(L_1)$. The quantity
$v\rho_{\rm in,eff}$ corresponds to the rate
with which particles attempt to enter the ASEP at its left boundary.
Note that (i) this effective boundary density depends on the current
$J$ and (ii) that $\rho_{\rm in,eff}$ can be larger than one.  The
right boundary density is given by $\rho_{\rm ex}$.

The phase diagram can now be determined in a self-consistent way.  As
in the tube system studied in Ref.\ \cite{Klumpp_Lipowsky2003}, the
same phases are found as for the one-dimensional ASEP, but the
location of the transition lines depends on the values of the model
parameters $v/D$ and $L_1$, and the maximal current phase may be shifted
out of the physically accessible range of the parameters.

The system is in the maximal current phase if $\rho_{\rm ex}<1/2$ and
$\rho_{\rm in,eff}>1/2$. In this case the current is $J=v/4$, and the condition
\begin{equation}
  \rho_{\rm in,eff}=\frac{D}{v}\rho_{\rm in}-\frac{L_1}{4}\geq\frac{1}{2}
\end{equation}
implies that the system is in the maximal current phase for
\begin{equation}
  \rho_{\rm in}\geq \frac{v}{2D}\left(1+\frac{L_1}{2}\right).
\end{equation}
For large $v/D$, the latter value of the left boudary density is
larger than one and therefore not physically accessible. This implies
that a maximal current phase is only present for small velocities with
$v/D<2/(1+L_1/2)$. If the velocity is larger, unphysically high
densities would be necessary at the left boundary in order to
establish a sufficiently large density gradient which could generate a
diffusive current with the value $v/4$, the maximal current defined by
the driven compartment.  In this situation, the diffusive compartments
acts as a diffusive bottleneck: If the maximally possible diffusive
current through the diffusive compartment is smaller than $v/4$, a
maximal current phase cannot occur, because the diffusive compartment
cannot maintain the maximal current.

A simpler estimate comparing the maximal diffusive current $D/L_1$,
which is restricted by the maximal density difference of one, with the
maximal driven current $v/4$ yields the condition, $v/D<4/L_1$, which
agrees with the previous one for large $L_1$, but is less restrictive
for small $L_1$. The latter discrepancy reflects the fact that the
maximal density difference in the diffusive compartment is actually
smaller than one since the density at $x=L_1$ must be larger than zero.

In addition, a low density phase is found for $\rho_{\rm in,eff}<1/2$
and $\rho_{\rm in,eff}<1-\rho_{\rm ex}$ and a high density phase for
$\rho_{\rm in,eff}>1-\rho_{\rm ex}$ and $\rho_{\rm ex}>1/2$.  Along
the transition line between the high density and low density phases we
can use $J=v\rho_{\rm ex}(1-\rho_{\rm ex})$ and obtain
\begin{equation}
  \rho_{\rm in}=\frac{v}{D}\left[ 1+(L_1-1)\rho_{\rm ex}-L_1\rho_{\rm ex}^2 \right]
\end{equation}
for the transition line between the low density and the high density
phase.  This line extends from the right upper corner of the phase
diagram to the right upper corner of the maximal current phase region.
If there is no maximal current phase, it ends at a point with
$\rho_{\rm in}=1$ and $\rho_{\rm ex}>1/2$.

\begin{figure}[h]
  \begin{center}
    \leavevmode
    \includegraphics[angle=-90,width=.9\textwidth]{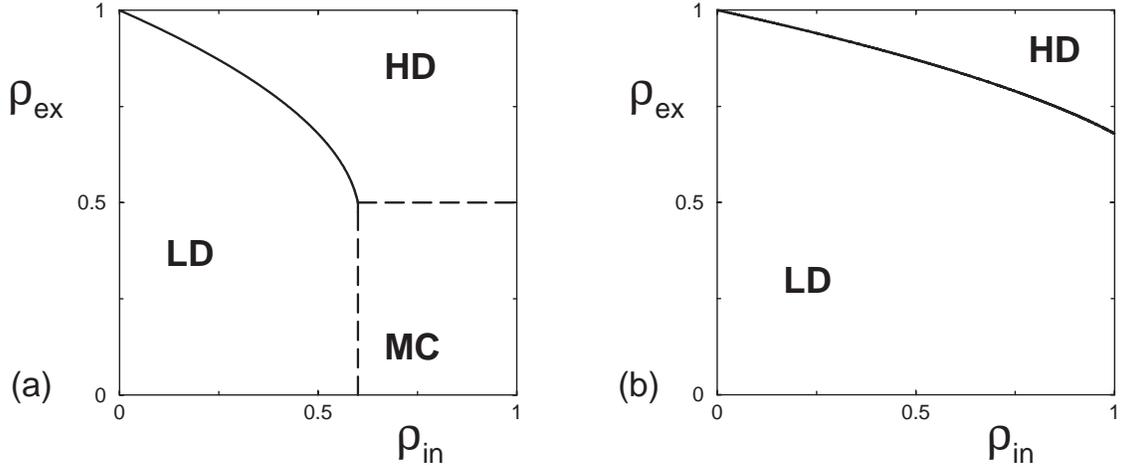}
    \caption{Phase diagrams for the ASEP with diffusive injection of 
      particles at the left boundary, i.e.\ for case (A) as shown in
      \fig{fig:Bildchen_asep_diff_inj_extr}, as a function of the left
      and right boundary densities $\rho_{\rm in}$ and $\rho_{\rm
        ex}$. The parameters are (a) $v=0.1$, $D=1/2$, and $L_1=10$; and 
      (b) $v=0.2$,
      $D=1/2$, and $L_1=10$.  }
    \label{fig:phasendiag_diff_inj}
  \end{center}
\end{figure}

Phase diagrams for two cases are shown in
\fig{fig:phasendiag_diff_inj}.  We have chosen $L_1=10$ and $D=1/2$ in
both cases.  The condition for the presence of a maximal current phase
is then $v<1/6$. In \fig{fig:phasendiag_diff_inj}(a), the velocity is
$v=0.1<1/6$ and all three phases are present, while in
\fig{fig:phasendiag_diff_inj}(b), $v=0.2>1/6$ and the maximal current
phase is absent. In the latter case the largest part of the phase
diagram is covered by the low density phase.

In the maximal current phase the current is $J=v/4$. In the high
density phase, it is determined by the right boundary density $\rho_{\rm
  ex}$ and has the value $J=v\rho_{\rm ex}(1-\rho_{\rm ex})$. Finally
in the low density phase, the current is given by the self-consistency
condition
\begin{equation}
  J=v\rho_{\rm in,eff}(J) [1-\rho_{\rm in,eff}(J)],
\end{equation}
which leads to a quadratic equation for the current.
The solutions is uniquely determined by the limits $J=0$ for
$\rho_{\rm in}=0$ and $J=v/4$ for $\rho_{\rm
  in}=\frac{v}{2D}(1+L_1/2)$ and is given by
\begin{equation}\label{strom_LD}
  J=\frac{v}{2L_1^2}\left(-1-L_1+2\frac{D}{v}L_1\rho_{\rm in} + \sqrt{\left(1+L_1-2\frac{D}{v}L_1\rho_{\rm in}\right)^2+4\frac{D}{v}L_1^2\rho_{\rm in}\left(1-\frac{D}{v}\rho_{\rm in}\right)} \right).
\end{equation}

\subsection{Diffusive extraction of particles}
\label{sec:diff_extr}

Next we consider case (B), in which particles which reach the end of
the active compartment have to diffuse over a distance $L_2$ before
they can leave the system at the right end.  This case can be treated
in the same way as the one with diffusive injection.  Note, however,
that it cannot simply be derived from the the latter one using
particle-hole symmetry, because a particle at the site left of the
driven compartment attempts to enter it with rate $D$, while a hole at
the site right of the driven compartment does so with rate $v$.

The density profile in the diffusive compartment is given by
$\rho(x)=\rho_{\rm ex}+(L+1-x)J/D$ for $L_1+1\leq x\leq L_1+L_2=L$.
Therefore the effective rate with which particles attempt to leave the
active compartment at $x=L_1$ is 
$v(1-\rho_{\rm ex}-L_2J/D)\equiv v(1-\rho_{\rm ex,eff})$ corresponding
to an effective right boundary density of the driven compartment given
by
\begin{equation} 
  \rho_{\rm ex,eff}=\rho_{\rm ex}+L_2J/D.
\end{equation}

The maximal current phase is now found for $\rho_{\rm in}>1/2$ and
\begin{equation}\label{eq:MCbeiDiffExtr}
  \rho_{\rm ex}<\frac{1}{2}-\frac{L_2 v}{4 D},
\end{equation}
which is always $\leq 1/2$. Again the maximal current phase is only
present if the range of boundary densities defined by
\eq{eq:MCbeiDiffExtr} is physically accessible. Here the corresponding
condition is $\rho_{\rm ex}>0$, which is valid for $v/D<2/L_2$.  The
latter condition expresses again the fact that the diffusive
compartment must also support this maximal current. The diffusive
current is, however, restricted by the maximally possible value of the
density gradient in the right (diffusive) compartment, $1/(2L_2)$,
which leads to a maximal diffusive current of $D/(2L_2)$. If the
latter current is smaller than $v/4$, a stationary maximal current
phase is absent; therefore, the presence of the maximal current phase
requires that the maximal diffusive current is $\geq v/4$ which leads
to the condition $v/D<2/L_2$.

The condition $1-\rho_{\rm ex,eff}=\rho_{\rm in}$ with $J=v\rho_{\rm
  in}(1-\rho_{\rm in})$ yields the transition line between the low
density and the high density phase
\begin{equation}
  \rho_{\rm ex}=1-\rho_{\rm in}(1+\frac{v}{D}L_2)+\frac{v}{D}L_2\rho_{\rm in}^2.
\end{equation}
For velocities larger than $2D/L_2$, which is the maximal value for
the occurrence of a maximal current phase, this line ends at a point
in the phase diagram with $\rho_{\rm in}=0$ and $\rho_{\rm ex}<1/2$.
In this case the high density phase covers most of the phase diagram.

The current is $J=v/4$ in the maximal current phase and $J=v\rho_{\rm
  in}(1-\rho_{\rm in})$ in the low density phase. In the high density
phase, it is again given by a self-consistency condition $J=v\rho_{\rm
  ex,eff}(1-\rho_{\rm ex,eff})$, where $\rho_{\rm ex,eff}$ is a
function of $J$,
from which we obtain 
\begin{equation}\label{strom_HD}
  J=\frac{v}{2(L_2 v/D)^2}\left( -1+\frac{v}{D}L_2(1-2\rho_{\rm ex}) +\sqrt{1-2\frac{v}{D}L_2+4\frac{v}{D}L_2\rho_{\rm ex}+\left(\frac{v}{D}L_2\right)^2}\right).
\end{equation}

\subsection{Both diffusive injection and extraction of particles}
\label{sec:diff_inj&extr}

Now we consider case (C), i.e., we combine the two preceding cases.
Transport is now driven in the middle compartment for which we have
the effective boundary densities
\begin{equation}
  \rho_{\rm in, eff}  = \frac{D\rho_{\rm in}}{v}-\frac{L_1 J}{v}\qquad{\rm and}\qquad
  \rho_{\rm ex, eff}  = \rho_{\rm ex}+L_3 J/D.
\end{equation}
The phase boundary between the low density phase and the maximal
current phase is not affected by adding another compartment at the
right end, thus we can use the result from case (A). Likewise the
phase boundary between the high density phase and the maximal current
phase is unaffected by the left diffusive compartment, so that we can
use the result from case (B) upon substituting $L_2$ with $L_3$.  The
maximal current phase is therefore found for
\begin{equation}
  \rho_{\rm in}> \frac{v}{2D}\left(1+\frac{L_1}{2}\right) \quad {\rm and}\quad \rho_{\rm ex}<\frac{1}{2}-\frac{L_3v}{4D}.
\end{equation}
It is present if $v/D<2/(1+L_1/2)$ and $v/D<2/L_3$.  In addition, the
current is given by Eqs.\ (\ref{strom_LD}) and (\ref{strom_HD}) in the
low density and the high density phase, respectively.

\begin{figure}[h]
  \begin{center}
    \leavevmode
    \includegraphics[angle=-90,width=.5\textwidth]{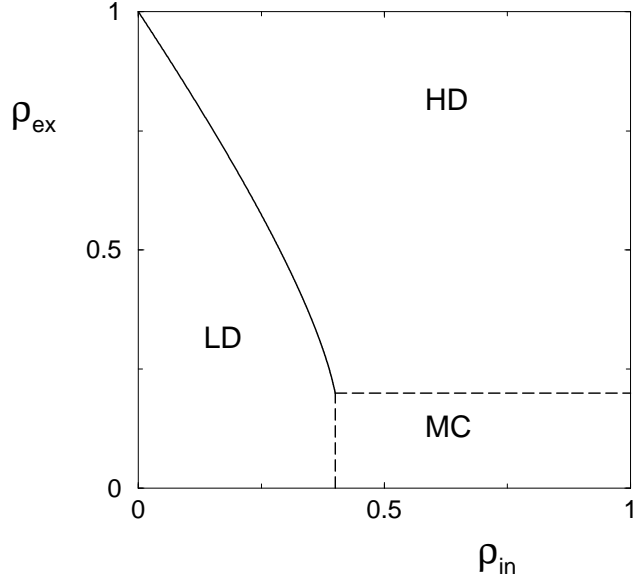}
    \caption{Phase diagram for the ASEP with diffusive injection and 
      extraction of particles, i.e. for the case (C) shown in
      \fig{fig:Bildchen_asep_diff_inj_extr}. The parameters are
      $v=0.1$, $D=1/2$, and $L_1=L_3=6$.}
    \label{fig:phasendiag_diff_inj_extr}
  \end{center}
\end{figure}

Finally, the transition line between these two phases is obtained from
the condition $\rho_{\rm in,eff}=1-\rho_{\rm ex, eff}$, which leads to
\begin{equation}
  \rho_{\rm in}=\frac{v}{D}\left[1-\rho_{\rm ex} + \frac{J(\rho_{\rm ex})}{v}(L_1-\frac{v}{D}L_3) \right],
\end{equation}
where $J(\rho_{\rm ex})$ is the current in the high density phase for
the right boundary density $\rho_{\rm ex}$ as given by \eq{strom_HD}.
The complete phase diagram is shown in
\fig{fig:phasendiag_diff_inj_extr}, where we have chosen parameters for
which a maximal current phase is present.

\subsection{Comparison with simulations}
\label{sec:simulations}

In addition to the self-consistency calculations presented above, we
performed Monte Carlo simulations. In this section, we compare the
simulation results with the mean field predictions for case (C).

\begin{figure}[h]
  \begin{center}
    \leavevmode
    \includegraphics[angle=-90,width=.5\textwidth]{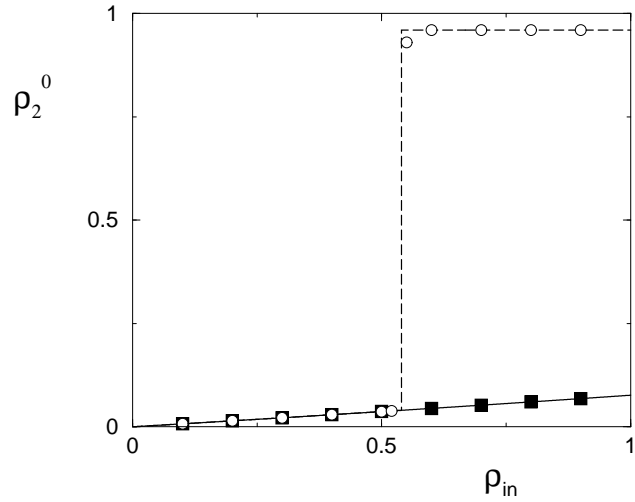}
    \caption{Bulk density $\rho_2^0$ in the active compartment of a system 
      with geometry (C) as a function of the left boundary density
      $\rho_{\rm in}$. Lines are the mean field results and symbols are
      simulation data. The right boundary density is fixed to
      $\rho_{\rm ex}=0$ (solid line) and $\rho_{\rm ex}=0.5$ (dashed
      line). Parameters have been chosen so that no maximal current
      phase is found; $v=1$, $D=1/2$, $L_1=L_3=6$ and $L_2=388$.}
    \label{fig:dichte_v1b6}
  \end{center}
\end{figure}

In the case where the maximal current phase is absent, i.e.\ for large
velocities, we find quantitative agreement of the measured current and
bulk density with the predictions of the mean field calculation. The
transition from the low density to the high density phase occurs at
the predicted density.  As an example, we show results for the bulk
density $\rho_2^0$ in the active compartments in Fig.\ 
\ref{fig:dichte_v1b6}, where we have chosen $\rho_{\rm ex}=0$ and
$\rho_{\rm ex}=0.5$. In the first case, the system is in the low
density phase for all values of $\rho_{\rm in}$, in the second case, a
transition to the high density phase is found at $\rho_{\rm in}\simeq
0.54$. The mean field results (lines) and the simulation data
(symbols) agree well.

\begin{figure}[h]
  \begin{center}
    \leavevmode
    \includegraphics[angle=-90,width=.5\textwidth]{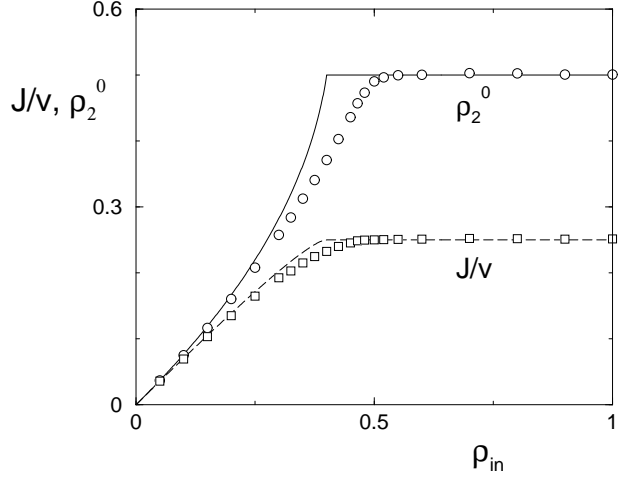}
    \caption{Current $J$ (dashed line) and bulk density $\rho_2^0$ 
      (solid line) in the active compartment of a system with geometry
      (C) as a function of the left boundary density $\rho_{\rm in}$
      for a case with transition to the maximal current phase. The
      lines are the predictions of the mean field calculation, the
      symbols simulation data. The parameters are $v=0.1$, $D=1/2$,
      $L_1=L_3=6$, $L_2=388$, and $\rho_{\rm ex}=0$.}
    \label{fig:strom_dichte_v.1b6}
  \end{center}
\end{figure}

If a maximal current phase is present, i.e.\ for small velocities, the
agreement is less good, although the phase diagram is still in
qualitative agreement with the mean field predictions.
\fig{fig:strom_dichte_v.1b6} shows again results for the case
$\rho_{\rm ex}=0$. Close to the transition to the maximal current
phase the current is smaller than predicted by the mean field
calculation.  Therefore the transition to the maximal current phase is
shifted towards a larger value of $\rho_{\rm in}$ and the increase of
the bulk density near the transition is less steep. Far from the
transition, however, agreement is again good. Likewise, the
transition line between the low density and the high density phase is also
shifted towards larger $\rho_{\rm in}$, as this transition line ends
on the phase boundary of the maximal current phase. Agreement becomes
again better far from the maximal current phase, since the other end
point of the line ($\rho_{\rm in}=0$, $\rho_{\rm ex}=1$) is exact.

\section{Diffusive bottleneck in the middle}
\label{sec:caseD}

Finally, let us consider case (D), where active transport is
interrupted by a diffusive compartment. In the case of molecular
motors, this can be realized by a gap in the filament network,
along which active transport takes place. Motors thus have to overcome
this gap by diffusion before they can rebind to a filament and
continue their active movements. If the middle compartment consists of
only one lattice site, $L_2=1$, this system reduces to the case of an
ASEP with a single defect which has been discussed previously, see
Refs.\ \cite{Janowsky_Lebowitz1992,Kolomeisky1998}.

The effective right boundary density for the left active compartment
is
\begin{equation}
  \rho_{\rm ex,eff}=\rho(L_1+1)
\end{equation}
and the effective left boundary density for the right active
compartment is
\begin{equation}
\rho_{\rm in,eff}=\frac{D\rho(L_1+L_2)}{v}=\frac{D\rho(L_1+1)-(L_2-1)J}{v}. 
\end{equation}

In this case, there are five possible phases. Because the current is
the same in both active compartments, the bulk densities in the left
and right compartment, $\rho_1^0$ and $\rho_3^0$, respectively, are
either equal or related by $\rho_1^0=1-\rho_3^0$. If the bulk
densities in both active compartments are equal, $\rho_1^0=\rho_3^0$,
there are three possibilities. Both compartments can be in the low
density, high density or maximal current phase. We denote these three
cases by LD--LD, HD--HD, and MC--MC, respectively. If the densities
are not equal, we have $\rho_1^0=1-\rho_3^0$, and there are two
additional possible phases, where one compartment is in the high
density and the other in the low density phase. These phases will be
denoted by HD--LD and LD--HD if $\rho_1^0$ is larger or smaller than
$\rho_3^0$, respectively.

\subsection{Phases with equal densities in the active compartments}

In the \emph{MC--MC phase}, the current is $J=v/4$ and we have four
conditions for the boundary densities, $\rho_{\rm in}>1/2$, $\rho_{\rm
  ex}<1/2$, $\rho_{\rm in,eff}>1/2$ and $\rho_{\rm ex,eff}<1/2$. The
first two conditions are the same as for an ASEP without the diffusive
compartment and the latter two conditions yield $v/D<2/(1+L_2)$. The
maximal current phase is found only for small velocities, for larger
velocities in the active compartments, the diffusive section acts
again as a diffusive bottleneck.

The \emph{LD--LD phase} is characterized by $J=v\rho_{\rm
  in}(1-\rho_{\rm in})=v\rho_{\rm in,eff}(1-\rho_{\rm in,eff})$. As
both $\rho_{\rm in}$ and $\rho_{\rm in,eff}$ must be smaller than
$1/2$, this implies $\rho_{\rm in}=\rho_{\rm in,eff}$.  Together with
the condition $\rho_{\rm ex}<1-\rho_{\rm in,eff}$ for the right active
compartment, this implies that the LD--LD phase is found within the
region of the phase diagram where the low density phase of the usual
ASEP is located.  An additional condition is given by $\rho_{\rm
  ex,eff}=\rho(L_1+1)<1-\rho_{\rm in}$. Thus, with the condition $\rho_{\rm
  in}=\rho_{\rm in,eff}$, we obtain $\rho(L_1+1)$ as a function of
$\rho_{\rm in}$,
\begin{equation}
  \rho(L_1+1)=\frac{v}{D}[\rho_{\rm in}+(L_2-1)\rho_{\rm in}(1-\rho_{\rm in})]
\end{equation}
which leads to the inequality
\begin{equation}\label{LDLD_UNGL}
  \frac{v}{D}[\rho_{\rm in}+(L_2-1)\rho_{\rm in}(1-\rho_{\rm in})]  +\rho_{\rm in}<1.
\end{equation}
Because $\rho_{\rm in}<1/2$, the left hand side of this inequality is
increasing monotonically and the solution is given by $\rho_{\rm
  in}<\rho_{\rm in,*}$, where $\rho_{\rm in,*}$ is the solution of the
equation which is obtained upon substituting '$<$' with '$=$' in Eq.\ 
(\ref{LDLD_UNGL}). This solution is uniquely determined by the
limiting case $L_2=1$, in which Eq.\ (\ref{LDLD_UNGL}) yields
$\rho_{\rm in}<1/(1+v/D)$.  As a result we obtain the condition
\begin{equation}\label{rho_in*}
  \rho_{\rm in}<\rho_{\rm in,*}\equiv\frac{\frac{v}{D} L_2+1-\sqrt{(\frac{v}{D} L_2+1)^2-4\frac{v}{D}(L_2-1)}}{2\frac{v}{D}(L_2-1)}.
\end{equation}
If $\rho_{\rm in,*}$ is larger than $1/2$, this condition does not
further restrict the occurrence of the LD--LD phase, since for
$\rho_{\rm in}=1/2$ the transition to the maximal current phase takes
place.  Indeed, $\rho_{\rm in,*}>1/2$ implies the simpler condition
$v/D<2/(L_2+1)$, which is exactly the condition we derived above as a
condition for the presence of the MC--MC phase.

On the other hand, if $\rho_{\rm in,*}<1/2$, condition
(\ref{LDLD_UNGL}) yields a restriction of the LD--LD phase to the
region in the phase diagram with $\rho_{\rm in}<\rho_{\rm in,*}$.
Therefore we conclude that for large velocities with $v/D>2/(L_2+1)$,
the transition to the MC--MC phase at $\rho_{\rm in}=1/2$ is replaced
by a transition to another phase at $\rho_{\rm in}=\rho_{\rm in,*}$.
The only possibility for this phase is the HD--LD phase which will be
discussed below.

Finally note that for $L_2=1$ our result recovers the condition for
the ASEP with a defect \cite{Kolomeisky1998}, where for $v/D<1$ the
phase diagram is the same as without the defect, while for $v/D>1$ a
phase dominated by the defect is found.

The \emph{HD--HD phase} can be treated in the same way and gives
corresponding results. A HD--HD phase can be found for $\rho_{\rm
  ex}>1-\rho_{\rm in}$ and $\rho_{\rm ex}>1/2$ for small velocities
which fulfill again the condition $v/D<2/(L_2+1)$, while for larger
velocities an additional restricting condition is found, namely
$\rho_{\rm ex}> \rho_{\rm ex,*}$ with $\rho_{\rm ex,*}\equiv
1-\rho_{\rm in,*}$.

\subsection{Phases with different densities in the active compartments}

\begin{figure}[h]
  \begin{center}
    \leavevmode
    \includegraphics[angle=-90,width=.5\textwidth]{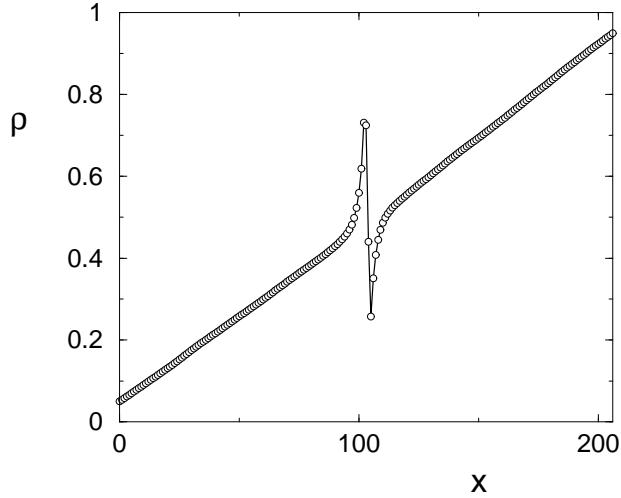}
    \caption{Density profile for the 'LD--HD' phase or coexistence of 
      the LD--LD and HD--HD phases for a system with a diffusive
      defect [case (D)]. As discussed in the text, the profile is essentially
      linear rather than consisting of two plateaus. The parameters
      are $v=1$, $D=1/2$, $L=205$, $L_1=L_3=101$, $L_2=3$, $\rho_{\rm
        in}=0.05$, and $\rho_{\rm ex}=0.95$.}
    \label{fig:profilLDHD}
  \end{center}
\end{figure}

For an \emph{LD--HD phase}, the current must be $J=v\rho_{\rm
  in}(1-\rho_{\rm in})=v\rho_{\rm ex}(1-\rho_{\rm ex})$. This implies
that the densities are $\rho_{\rm in}=1-\rho_{\rm ex}$, i.e., a
stationary state with a low density in the left, but a high density in
the right active compartment can only be expected along the line,
where the low density and high density phases coexist in the ASEP
without a diffusive compartment. In this case, however, a domain wall
diffuses through the system in the usual ASEP resulting in a density
profile which increases linearly \cite{Kolomeisky__Straley1998}. The
same can be expected for our case with the exception of the regions
close to the diffusive compartment. This behavior has previously been
observed in simulations for the case of an ASEP with a defect
\cite{Kolomeisky1998} which correspond to our case with $L_2=1$ and we
find the same behavior in Monte Carlo simulations for larger $L_2$,
see \fig{fig:profilLDHD}.

Finally, let us consider the possibility of a \emph{HD--LD phase}. In
this case the current is $J=v\rho_{\rm in,eff}(1-\rho_{\rm
  in,eff})=v\rho_{\rm ex,eff}(1-\rho_{\rm ex,eff})$ which, together with the conditions 
$\rho_{\rm in,eff}<1/2$ and $\rho_{\rm ex,eff}>1/2$, implies
$\rho_{\rm in,eff}=1-\rho_{\rm ex,eff}$. 
Substituting the
expression for the effective boundary densities, we obtain a quadratic
equation for $\rho(L_1+1)$ with the solution 
\begin{equation}
  \rho(L_1+1)=\rho_{\rm ex,*}=1-\rho_{\rm in,*}
\end{equation}
with $\rho_{\rm in,*}$ as given by Eq.\ 
(\ref{rho_in*}). For $L_2=1$ we recover again the corresponding result
for the ASEP with a defect.

\subsection{Phase diagrams}

\begin{figure}[h]
  \begin{center}
    \leavevmode
    \includegraphics[angle=-90,width=.9\textwidth]{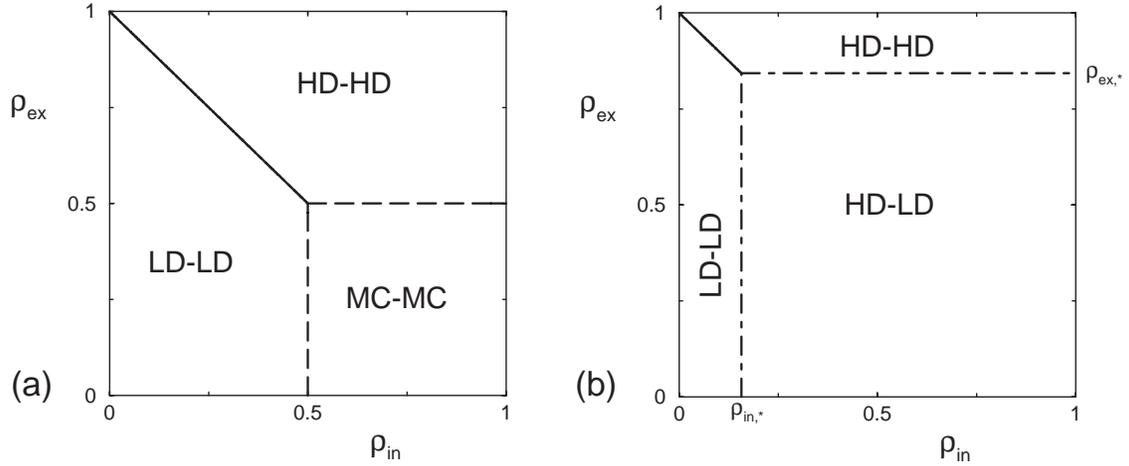}
    \caption{Phase diagrams for case (D), i.e.\ for a diffusive compartment 
      between two active compartments, as a function of the left and
      right boundary density, $\rho_{\rm in}$ and $\rho_{\rm ex}$. (a)
      Small velocity, $v=0.1$ and (b) large velocity, $v=1$. The
      diffusive section has length $L_2=3$ and the diffusion
      coefficient is $D=1/2$. }
    \label{fig:phasendiag_diffZwischenstueck}
  \end{center}
\end{figure}

We can now summarize the results into phase diagrams as shown
in \fig{fig:phasendiag_diffZwischenstueck}. There are two different
cases: If $v/D<2/(1+L_2)$, the phase diagram corresponds to the one of
the ASEP without the diffusive compartment, see
\fig{fig:phasendiag_diffZwischenstueck}(a). The bulk density is equal
in both active compartments. If, on the other hand, the velocity is
larger with $v/D>2/(1+L_2)$, the diffusive compartment acts as a
bottleneck and the phase diagram is modified, see
\fig{fig:phasendiag_diffZwischenstueck}(b). We find the LD--LD phase
for $\rho_{\rm in}<\rho_{\rm in,*}$ and $\rho_{\rm in}<1-\rho_{\rm
  ex}$ and the HD--HD phase for $\rho_{\rm ex}>\rho_{\rm ex,*}$ and
$\rho_{\rm ex}>1-\rho_{\rm in}$. For $\rho_{\rm in}>\rho_{\rm in,*}$
and $\rho_{\rm ex}<\rho_{\rm ex,*}$, the system exhibits the HD--LD
phase. While in the LD--LD and HD--HD phases, the bulk densities in
the active compartments and the current are determined by the boundary
densities, these quantities are independent of the boundary densities
in the HD--LD phase and depend only on the ratio $v/D$ and the length
$L_2$ of the diffusive compartment.

\begin{figure}[h]
  \begin{center}
    \leavevmode
    \includegraphics[angle=-90,width=.5\textwidth]{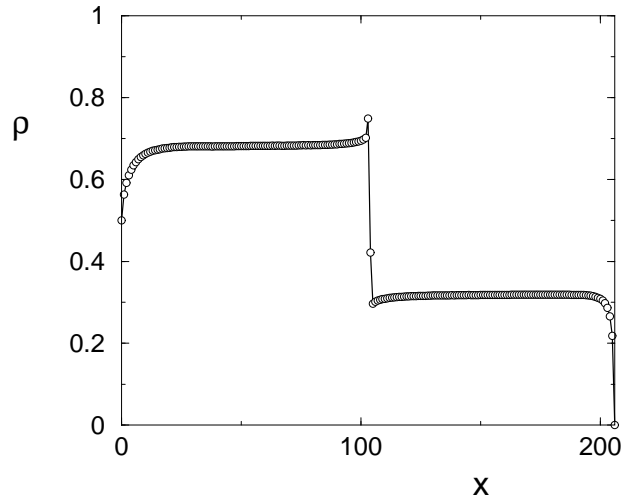}
    \caption{Density profile for the HD--LD phase in case (D) with 
      $\rho_{\rm in}=0.5$, $\rho_{\rm ex}=0$, $v=0.25$, and $D=1/2$.
      The geometrical parameters are as in \fig{fig:profilLDHD}. }
    \label{fig:profilHDLD}
  \end{center}
\end{figure}

The HD--LD phase has some similarities to a maximal current phase. The
current is constant throughout this phase and attains its maximal
value compared to all other phases, $J=v\rho_{\rm in,*}(1-\rho_{\rm
  in,*})=v\rho_{\rm ex,*}(1-\rho_{\rm ex,*})$. This maximal value,
however, is not determined by the active compartments, but corresponds
to the maximal current which can be supported by the diffusive
compartment. Correspondingly, the density profiles in the HD--LD phase
(shown in \fig{fig:profilHDLD}) do not exhibit the power-law behavior
known from the usual maximal current phase. The bulk densities in the
left and right active compartments are constant in this phase as well
and are given by $\rho_{\rm ex,*}$ and $\rho_{\rm in,*}$,
respectively.  Note, however, the following difference compared to the
usual maximal current phase. The transitions to the maximal current
phase in the usual ASEP
are continuous. 
The transitions to the HD--LD phase in our case are somewhat peculiar
in the sense that they are continuous in one compartment, but
discontinuous in the other. For example, at the transition from the
LD--LD phase to the HD--LD phase, the density $\rho_3^0$ in the right
active compartment is continuous, but the bulk density $\rho_1^0$ in
the left active compartment exhibits a jump from $\rho_{\rm in,*}$ to
$1-\rho_{\rm in,*}$.

We performed again Monte Carlo simulations and compared the results
with the mean field predictions. As a result, we find that the phase
diagram obtained by the self-consistent mean field approach is
recovered for small velocities. For large velocities, on the other
hand, the qualitative behavior is correctly predicted by the mean
field calculation, but the transition lines for the transition to the
HD--LD phase are shifted.  For example for $v=1$, $D=1/2$, and
$L_2=3$, we find the transition from the LD--LD phase to the HD--LD
phase at $\rho_{\rm in}\simeq 0.08$ in the simulations, while our mean
field calculation predicts a transition at $\rho_{\rm in}=\rho_{\rm
  in,*}\simeq 0.13$.  Correspondingly, there are also differences in
the values for the current in the HD--LD phase and the critical value
of the velocity is found to be smaller than predicted by the mean
field calculation. For $L_2=3$ and $D=1/2$, we observe the usual 
maximal current
phase for $v\protect\siml 0.15$ in the simulations while the mean
field approach yields $v<0.25$.

Finally, let us add a remark concerning the density profiles in the
HD--LD phase. From our mean field approach, we expect the constant
bulk densities in the left and right active compartments to be
approached exponentially from the left and right boundary,
respectively. This is the case in the simulation data; however, in
addition, an excess density close to the diffusive compartment is
observed, which is not expected from the mean field approach, see
\fig{fig:profilHDLD}. As reported previously for the case of an ASEP
with a single defect site \cite{Janowsky_Lebowitz1992}, this excess
density decays as a power-law $\sim x^{-1}$, so that some kind of
long-range order is also present in this phase which plays the role of
a maximal current phase for the system with a diffusive bottleneck.

\section{Summary and conclusions}

We have discussed transport in one-dimensional lattices which consist
of two or three compartments where transport is either driven or
diffusive, a situation that is inspired by the motion of molecular
motors which diffuse until they reach a filament and then move along
that filament in a directed manner \cite{Klumpp_Lipowsky2003}. 
Mutual exclusion from lattice
sites is important, as many particles can be injected into the system
from reservoirs of fixed density at both ends. This is again realistic
for molecular motors, which are strongly attracted by the filaments,
so that the density of motors along the filaments can become large,
even if the motor concentration in solution is relatively small.
Traffic in the compartments where transport is active or driven is
described by an asymmetric simple exclusion process.

We have studied four different geometries. In the cases (A)--(C),
particles are injected into an active compartment and/or extracted
from it via diffusive compartments. In case (D), active transport is
interrupted by a diffusive compartment in the middle. The latter case
is a generalization of the ASEP with a point defect.

In all cases, the diffusive compartments can act as diffusive
bottlenecks.  If the velocity of the particles in the driven
compartment is sufficiently small, the phase diagram is essentially
the same as for the usual one-dimensional ASEP. In the cases (A)--(C),
the locations of the transition lines depend on the model parameters.
For large velocities, on the other hand, transport through the lattice
is limited by the diffusive compartments which cannot support
arbitrarily large currents.  In the cases (A)--(C), this situation
implies the absence of the maximal current phase.

In contrast, in case (D), this leads to a peculiar new phase, the
HD--LD phase, where the density is high in the left, but low in the
right active compartment. As in a maximal current phase, the current
is constant in the HD--LD phase and has the maximal possible value. In
contrast to the usual maximal current phase, this value of the currrent 
is determined
by the diffusive compartment, and the transitions to the HD--LD phase
are continuous in one, but discontinuous in the other active
compartment.

%
%

\newpage

\end{document}